\documentclass[%
 aip,apl,
 amsmath,amssymb,
 reprint,%
]{revtex4-1}

\usepackage{graphicx}% Include figure files
\usepackage[utf8]{inputenc}
\usepackage[T1]{fontenc}
\usepackage{etoolbox}
\usepackage{lineno}
\usepackage{xcolor}
\usepackage{comment}

\newcommand{\SAP}{Dipartimento di Fisica - Sapienza Universit\`{a} di Roma, Piazzale Aldo Moro 2, 00185, Roma - Italy}
\newcommand{\INFNRM}{INFN - Sezione di Roma, Piazzale Aldo Moro 2, 00185, Roma - Italy}
\newcommand{\INFNFE}{INFN - Sezione di Ferrara, via Saragat,1 44121, Ferrara - Italy}
\newcommand{\UNIFE}{Dipartimento di Fisica e Scienze della Terra, Università di Ferrara, Via Saragat 1, 44100, Ferrara, Italy}
\newcommand{\UNIFEN}{Dipartimento di Neuroscienze e Riabilitazione, Universit\`{a} di Ferrara, Via Luigi Borsari 46, 44121 Ferrara, Italy}
\newcommand{\NEEL}{Univ. Grenoble Alpes, CNRS, Grenoble INP, Institut N\'eel, 38000 Grenoble, France}
\newcommand{\CNRNANO}{Istituto di Nanotecnologia - CNR,  Piazzale Aldo Moro 2, 00185, Roma - Italy}
\newcommand{\CNRPHOTO}{Istituto di Fotonica e Nanotecnologie - CNR, Via del Fosso del Cavaliere 100, I-00133 Roma, Italy}

\makeatletter
\def\@email#1#2{%
 \endgroup
 \patchcmd{\titleblock@produce}
  {\frontmatter@RRAPformat}
  {\frontmatter@RRAPformat{\produce@RRAP{*#1\href{mailto:#2}{#2}}}\frontmatter@RRAPformat}
  {}{}
}%
\makeatother
\begin{document}
%\linenumbers

\title{BULLKID: Monolithic array of particle absorbers sensed by Kinetic Inductance Detectors}

\author{A.~Cruciani}\affiliation{\INFNRM}
\author{L.~Bandiera}\affiliation{\INFNFE}
\author{M.~Calvo}\affiliation{\NEEL}
\author{N.~Casali}\affiliation{\INFNRM}
\author{I.~Colantoni}\affiliation{\CNRNANO}\affiliation{\INFNRM}
\author{G.~Del~Castello}\affiliation{\SAP}\affiliation{\INFNRM}
\author{M.~del~Gallo~Roccagiovine}\affiliation{\SAP}\affiliation{\INFNRM}
\author{D.~Delicato}\affiliation{\SAP}\affiliation{\INFNRM}
%\author{S.~Di Domizio}
%\affiliation{Dipartimento di Fisica - Universit\`{a} degli Studi di Genova, Via Dodecaneso 33, 16146, Genova - Italy}
%\affiliation{INFN - Sezione di Genova, Via Dodecaneso 33, 16146, Genova - Italy}
%\author{J.~Zhou}
%\affiliation{Dipartimento di Fisica - Sapienza Universit\`{a} di Roma, Piazzale Aldo Moro 2, 00185, Roma - Italy}
\author{M.~Giammei}\affiliation{\SAP}\affiliation{\INFNRM}
\author{V.~Guidi}\affiliation{\UNIFE}\affiliation{\INFNFE}
\author{J.~Goupy}\affiliation{\NEEL}
\author{V.~Pettinacci}\affiliation{\INFNRM}
\author{G.~Pettinari}\affiliation{\CNRPHOTO}
\author{M.~Romagnoni}\affiliation{\INFNFE}
\author{M.~Tamisari}\affiliation{\UNIFEN}\affiliation{\INFNFE}
\author{A.~Mazzolari}\affiliation{\INFNFE}
\author{A.~Monfardini}\affiliation{\NEEL}
\author{M.~Vignati}\email{marco.vignati@roma1.infn.it}\affiliation{\SAP}\affiliation{\INFNRM}

\date{\today}% It is always \today, today,

\begin{abstract}
We introduce BULLKID, an innovative phonon detector consisting of an array of  dices acting as particle absorbers sensed by multiplexed Kinetic Inductance Detectors (KIDs). 
The dices are carved in a thick crystalline wafer and form a monolithic structure. The carvings leave a thin common disk intact in the wafer,  acting both as holder for the dices and as substrate for the KID lithography.
The  prototype  presented consists of an array of 64  dices of 5.4x5.4x5~mm$^3$ carved in a 3" diameter, 5 mm thick  silicon wafer, 
with a common disk 0.5 mm thick hosting a 60 nm patterned aluminum layer. The resulting array is highly segmented but avoids the use of dedicated holding structures for  each unit.
Despite the fact that the  uniformity of the KID electrical response across the array needs optimization, the operation of 8 units with similar  features shows, on average, a baseline energy resolution of $26\pm7$~eV.
This makes it a suitable detector for low-energy processes such as direct interactions of dark matter and  coherent  elastic neutrino-nucleus scattering. 

\end{abstract}

\maketitle

%%%%%%%%%%%%%%%%%%%%%%%%%%%%%%%%%%%%%%%%%%%%%%%
%intro
%%%%%%%%%%%%%%%%%%%%%%%%%%%%%%%%%%%%%%%%%%%%%%%

The  low energy threshold and the high energy resolution that cryogenic phonon detectors can achieve 
make them a competitive choice in particle-physics experiments searching for weak or rare signals, 
such as the interaction of dark matter particles with ordinary matter~\cite{SuperCDMS:2017mbc,CRESST2019,EDELWEISS:2016boq} or
the neutrino coherent and elastic scattering off atomic nuclei~\cite{Strauss:2017cuu,MINER}.
In these experiments the detector unit consists of a target crystal acting as particle absorber, where the energy released is converted to phonons.
By using  semiconducting~\cite{RICOCHETAG} or superconducting~\cite{SuperCDMSCPD,StraussGram} thermometers as phonon sensors, 
energy resolutions from  few electronvolts to tens of electronvolts have been achieved recently using crystals of mass between fraction of grams
to tens of grams.

Several experiments are aiming at target masses of kilograms but at the same time a high detector segmentation should be maintained for the identification of background processes. 
With absorber masses of grams or fraction of grams, arrays of hundreds or thousands of detector units will be needed, representing one of the main challenges in the field.
Phonon  sensors, indeed, require the operation in dilution refrigerators and this is a complication for the readout and for the assembly of large arrays. 
Nonetheless,  the structure holding the crystals is usually inert and this can be an issue in terms of background events due to radioactive contaminations~\cite{STRAUSS2017414}.

We propose a new  concept of array of phonon detectors, dubbed BULLKID, which is at the same time highly segmented, fully active and which can be easily fabricated. 
The  concept consists in carving a square grid in one of the two surfaces of a ``thick'' wafer. 
The carvings stop near  the opposite surface in order to create volumes with the shape of a dice acting as  particle absorbers. 
The opposite surface of the wafer is left intact, it forms a disk common to all dices and  hosts the sensors, each coupled to a dice.
We choose Kinetic Inductance Detectors (KIDs) as sensors,  
because of their built-in multiplexing capability and relatively easy fabrication.

KIDs~\cite{Day:2003fk} are superconducting resonators sensitive to the variation of the Cooper pair density. 
When energy is absorbed in the superconductor it can break the pairs and vary their kinetic inductance, giving rise to measurable changes in the frequency of the resonator.  
KIDs  are used as direct radiation absorbers in millimeter wavelength astronomy~\cite{refId0} and in single photon low-resolution spectrometers at visible to near-infrared wavelengths~\cite{Walter_2020}. 

The detection with KIDs of athermal phonons generated by interactions in an absorber was proposed a decade ago in order to increase the active volume~\cite{swenson,moore2}  and today is object of  R\&D~\cite{Chang:2018vn,wifikid}.  
So far the best energy resolutions have been obtained by the CALDER project~\cite{CalderWhitePaper}, which created devices consisting of a single KID of 2x2~mm$^2$ active area deposited on a silicon absorber $380~\mu$m thick and  2x2~cm$^2$ wide. 
The energy resolution is $80$ or $26$~eV, using as superconductors aluminum~\cite{Cardani:2017qr} or an aluminum-titanium-aluminum tri-layer~\cite{Cardani_2018}, respectively. The energy conversion efficiency, which is driven by the fraction of phonons reaching the KID and breaking  Cooper pairs, is key to reach good energy resolutions in these devices.  
The value obtained in CALDER, which we use as reference for this work, is  $\eta\sim10\%$~\cite{Cardani:2017qr,martinez2019} . 

%%%%%%%%%%%%%%%%%%%%%%%%%%%%%%%%%%%%%%%%%%%%%%%
%detector
%%%%%%%%%%%%%%%%%%%%%%%%%%%%%%%%%%%%%%%%%%%%%%%
In this Letter we present the first BULLKID prototype that demonstrates the feasibility and the potential of the approach. 
We describe the production,  the operation and  the performance  with focus on the phonon signal characterization  and on the energy resolution measurement.
%, in view of future improvements and applications.

\begin{figure}[t]
\begin{center}
\includegraphics[width=\columnwidth]{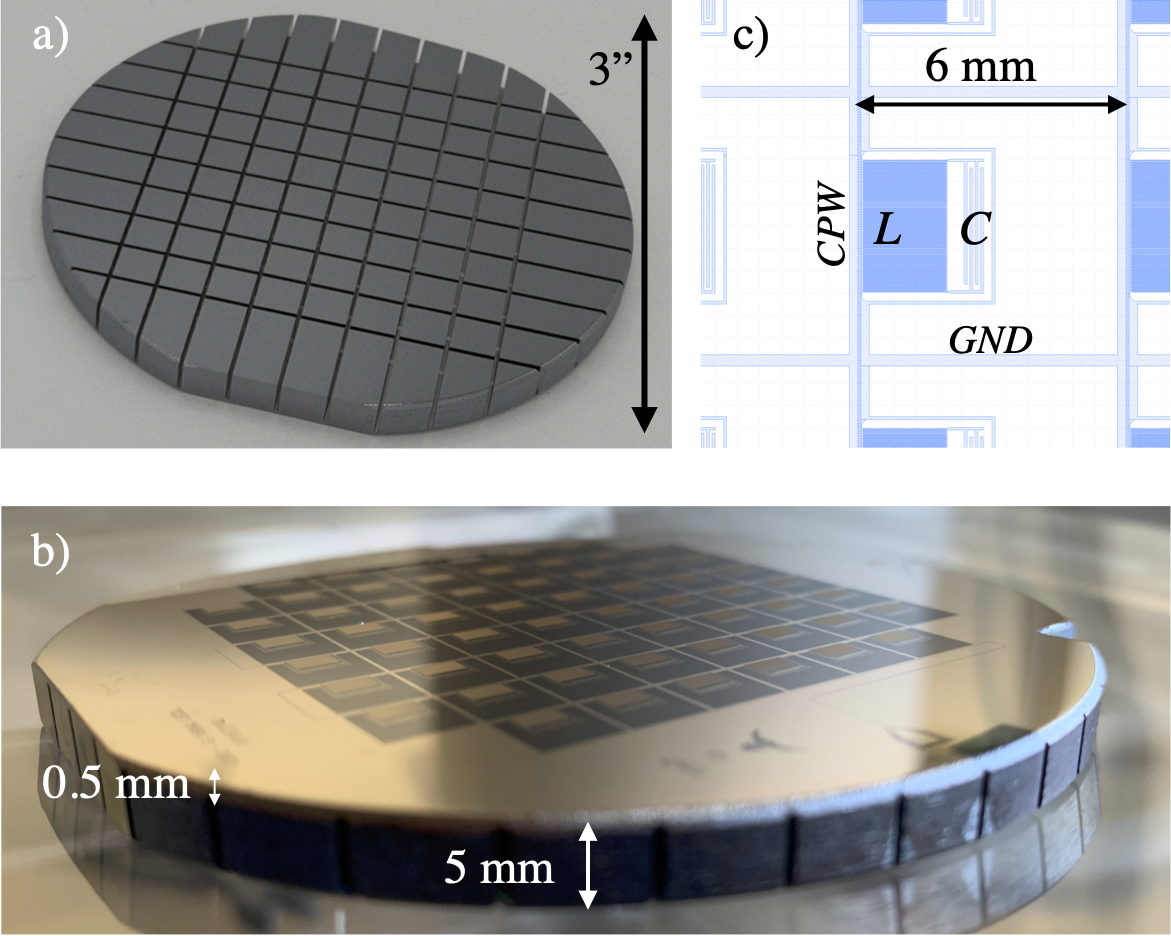}
\caption{
{\bf a)} Carved silicon wafer of 3" diameter and 5~mm thickness. Carvings are 4.5~mm deep and create 64 dices of 5.4x5.4x5~mm$^3$ each, leaving intact a 0.5~mm thick common disk. 
{\bf b)} Front side of the wafer with the multiplexed array of 60 aluminum KIDs fabricated on the common disk, with each KID sensing phonons from a single dice.  
{\bf c)} Layout of a  KID. A meandered inductor $L$  is closed on a capacitor $C$ and coupled to a coplanar waveguide (CPW) that runs through the array. The resonant frequency of each KID is set by adjusting individually the length of the capacitor's fingers.
 }
\label{fig1}
\end{center}
\end{figure}

The prototype  is created starting from a 3" diameter, 5 mm thick, (100) intrinsic and high-resistivity silicon wafer. The substrate is coated with a silicon nitride layer through low chemical vapor deposition. The carvings, performed with a circular saw dicing machine, are 0.6 mm wide and 4.5 mm deep. {The common disk is 0.5 mm thick}. The grid has 8x8 elements and a pitch of 6 mm, so that 64 dices of 5.4x5.4x5 mm$^3$ (0.34~g) are obtained~(Fig.~\ref{fig1}a). In order to obtain a robust structure, the orientation of the grid is chosen to let the carvings run in directions away from the crystallographic planes of the silicon. 
To prevent the trapping of phonons in the crystal defects created by the carvings,  the wafer is etched in an acid solution~\cite{Schwartz_1976}. In this step the silicon nitride layer acts as protective agent against the chemical etching~\cite{Baricordi_2008}. In the end the silicon nitride layer is removed via hydrofluoric acid wet etching.

The  wafer surface opposite to the carvings is instrumented with a multiplexed array of 60 aluminum KIDs, with each KID coupled to a dice. 
At the border a corona  $\sim1$~cm wide is not instrumented to allow handling, thus leaving  4 dices at the grid corners not sensed  (Fig.~\ref{fig1}b).
The KIDs have  an active area  of 4~mm$^2$ each, the same of CALDER~\cite{Cardani:2017qr} sensors for an easier comparison of the performances.
The layout  consists of a meandered inductor ($L$) 20~cm long and 20~$\mu$m wide, made of 106 segments 1.88~mm long interleaved by 8~$\mu$m.
The inductor is closed on a  interdigitated capacitor ($C$) with 4 fingers, whose length is adjusted to tune the resonator frequency $f_0 = 1 /2\pi \sqrt{LC}$  between 0.77 and 0.88 GHz (Fig.~\ref{fig1}c).
Each KID is inductively coupled to a single coplanar waveguide (CPW) running through the entire array. The lithography of the KID is made out of an aluminum film 60 nm thick
and is performed in a single step. The wafer is then clamped at the border by teflon supports in a 3D-printed copper holder and bonded to SMA connectors for the readout.

The wafer holder is anchored to the coldest point of a dry $^3$He/$^4$He dilution refrigerator with base temperature $T_0=20$~mK. 
In order to protect the KIDs  from thermal radiation and from magnetic fields, the detector is placed inside a shielding pot consisting of three layers of aluminum, copper and Cryophy\textsuperscript{\textregistered}~\cite{Cardani:2021wl}. 
For the energy calibration, 8 optical fibers individually excited at room temperature by a controlled 400~nm LED lamp face selected dices in the array, shining light on the side opposite to the KIDs.  
The RF bias of the KIDs is generated at room temperature with an Ettus X310 board~\cite{ettus} and fed in the cryostat where it is attenuated by $55$~dB before reaching the array. 
The output of the array is fed into a HEMT low-noise amplifier placed at the 4~K stage of the cryostat and then routed outside, back in the same board for the downconversion of the RF waves. 
For the data processing, an opensource firmware for the board is used~\cite{minutolo}, customized in order to trigger and acquire impulsive signals.

%%%%%%%%%%%%%%%%%%%%%%%%%%%%%%%%%%%%%%%%%%%%%%%
% frequency scans
%%%%%%%%%%%%%%%%%%%%%%%%%%%%%%%%%%%%%%%%%%%%%%%

\begin{figure}[t]
\begin{center}
\includegraphics[width=\columnwidth]{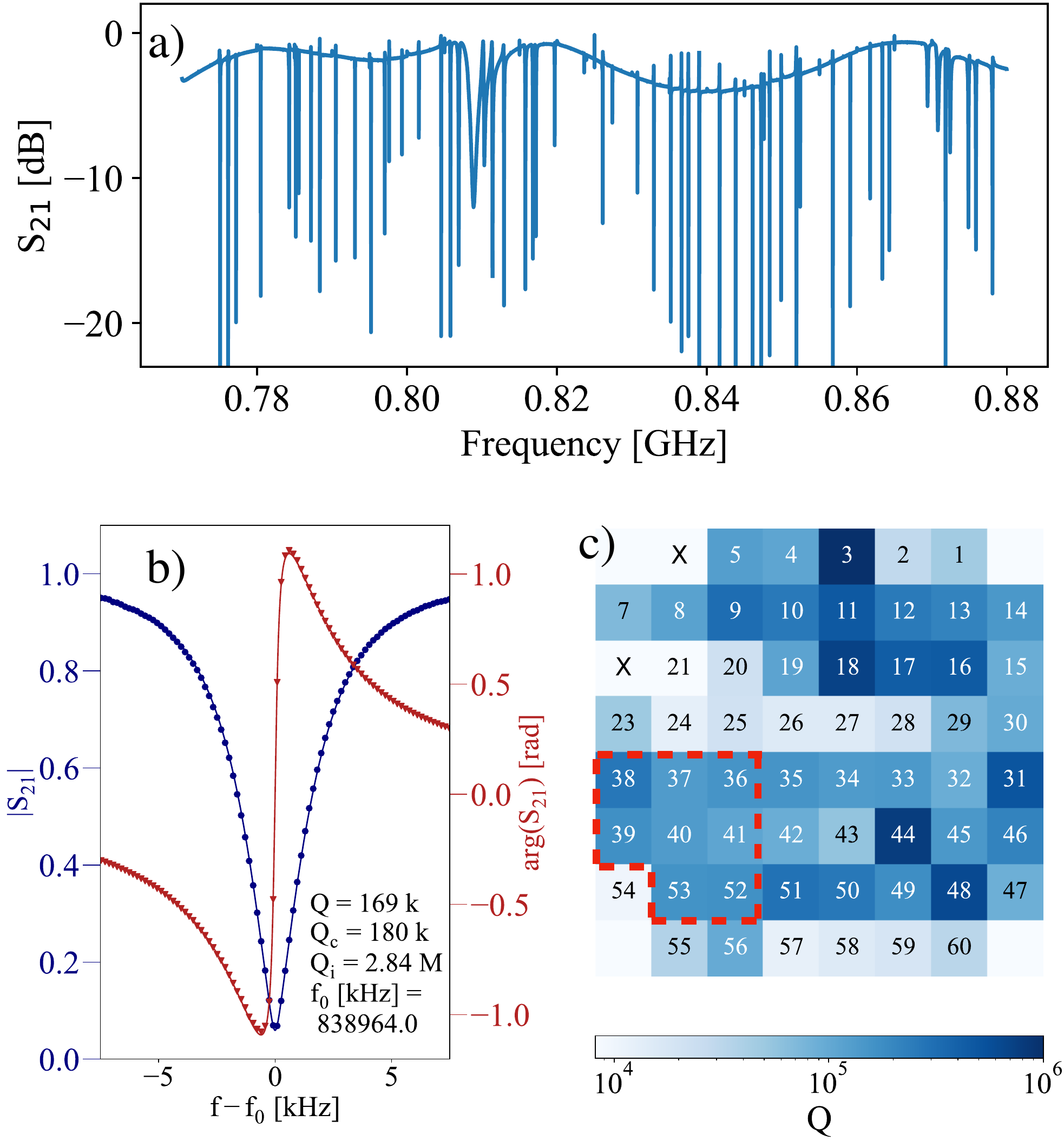}
\caption{
{\bf a)} Transmission $S_{21}$ as a function of the readout frequency spanning across the array. Each dip correspond to a KID and 58/60 of them are identified.
{\bf b)} Fit for Eq~\ref{eq:s21} on KID-40 (lines), using transmission data corrected for line losses and impedance mismatches (dots).
{\bf c)} Total quality factor $Q$ of each KID in the array evaluated from the resonance fitting. The KIDs in the red-dashed contour are used for the analysis of the phonon signal and of the energy resolution.
}
\label{fig2}
\end{center}
\end{figure}

The RF transmission past the device as a function of frequency, $S_{21}(f)$, is shown in Fig. 2a. 58 out of 60 (97\%) resonators are identified. 
 In order to extract the electrical features of each resonator the data are corrected offline for impedance mismatches and line losses~\cite{khalil} and then fitted 
 %simultaneously in the magnitude and phase planes 
with the model~\cite{zmu_annrev2012}:
\begin{equation}\label{eq:s21}
S_{21}(f) = 1 - \frac{Q}{Q_c}\frac{1}{1+j 2 Q \frac{f-f_0}{f_0}}
\end{equation}
where $Q^{-1}=Q_c^{-1} + Q_i^{-1}$ is the total quality factor,  $Q_c$ and $Q_i$ are the coupling and internal quality factors, respectively~(Fig.~\ref{fig2}b). 
By fitting  all the resonators we find $f_0$ in good agreement with the design and $Q_i\gg Q_c$, revealing the good quality of the film. The distribution of   $Q\simeq Q_c$ is however quite widespread around the design value of $1.5\cdot 10^5$, with some resonator $Q$ below  $10^4$ and some exceeding $5\cdot10^5$ (Fig~\ref{fig2}c). In principle $Q_c$ is determined by design via the distance between resonator and feed-line, which is precisely fixed by the lithography. 
However the propagation in the CPW of the slot mode alter the coupling at different positions along the line. Ways to correct for this issue consist in performing several bondings along the feedline  between the two ground planes, in order to fix their potential and to prevent the occurrence of the slot mode.

When an amount of  energy $E$ is released in a dice, changes of the Cooper pair density are detected by monitoring the time variation of $S_{21}$ with the RF bias set at $f=f_0$.  Changes in both the magnitude and phase of $S_{21}$ can be used, even tough the latter usually feature a better signal to noise ratio~\cite{zmu_annrev2012}. Restricting the analysis to the phase readout, its  variation with energy can be expressed in terms of the device parameters as:
\begin{equation}
\label{eq:response}
\frac{d\phi}{dE} = \eta\cdot \frac{\alpha S_2(f_0,T_0) Q}{N_0\Delta^2_0 V}
\end{equation}
where $N_0=1.72\cdot 10^{10}~{\rm eV}^{-1}\mu{\rm m}^{-3}$ is the single spin density of states in aluminum, $V = 2.4\cdot10^5~\mu{\rm m}^3$ is the inductor volume,
$S_2 (f_0, T_0)=4.9$ is a dimensionless factor given by the Mattis–Bardeen theory related the imaginary part of the conductivity~\cite{zmu_annrev2012}, $\alpha$ is the fraction of kinetic inductance with respect to the total inductance and $\Delta_0$ is the superconducting gap of the film. The values of $\alpha$ and $\Delta_0$ are estimated from the measured shift of the resonant frequency of each KID with temperature~\cite{GAOvsMattisBardeen}. By performing a scan from 20 to 350~mK, we find  values averaged over all KIDs of $\alpha = (5.0\pm0.1)$\% and $\Delta_0 = 189\pm 2~\mu$eV, where the errors are the standard deviations of the distributions.
The expected amplitude of the phase signal for $Q=1.5\cdot10^5$ is then $d\phi/dE \simeq 26\cdot \frac{\eta}{[0.1]} $~mrad/keV. This value is a factor 4 larger than in CALDER~\cite{Cardani:2017qr},  a factor 2  from the value of $S_2$  at $f_0\sim0.8$~GHz in place of $2.7$~GHz, and another factor 2  from the increase of $\alpha$, achieved with a modified design of the inductor, which is made of narrower segments. The actual value of $\eta$ is derived from the study of the  response to optical pulses, as detailed in the following.

%%%%%%%%% %%%%%%%%%%%%%%%%%%%%%%%%%%%%%%%%%%%%%%
%results 
%%%%%%%%%%%%%%%%%%%%%%%%%%%%%%%%%%%%%%%%%%%%%%%

\begin{figure}[t]
\begin{center}
\includegraphics[width=\columnwidth]{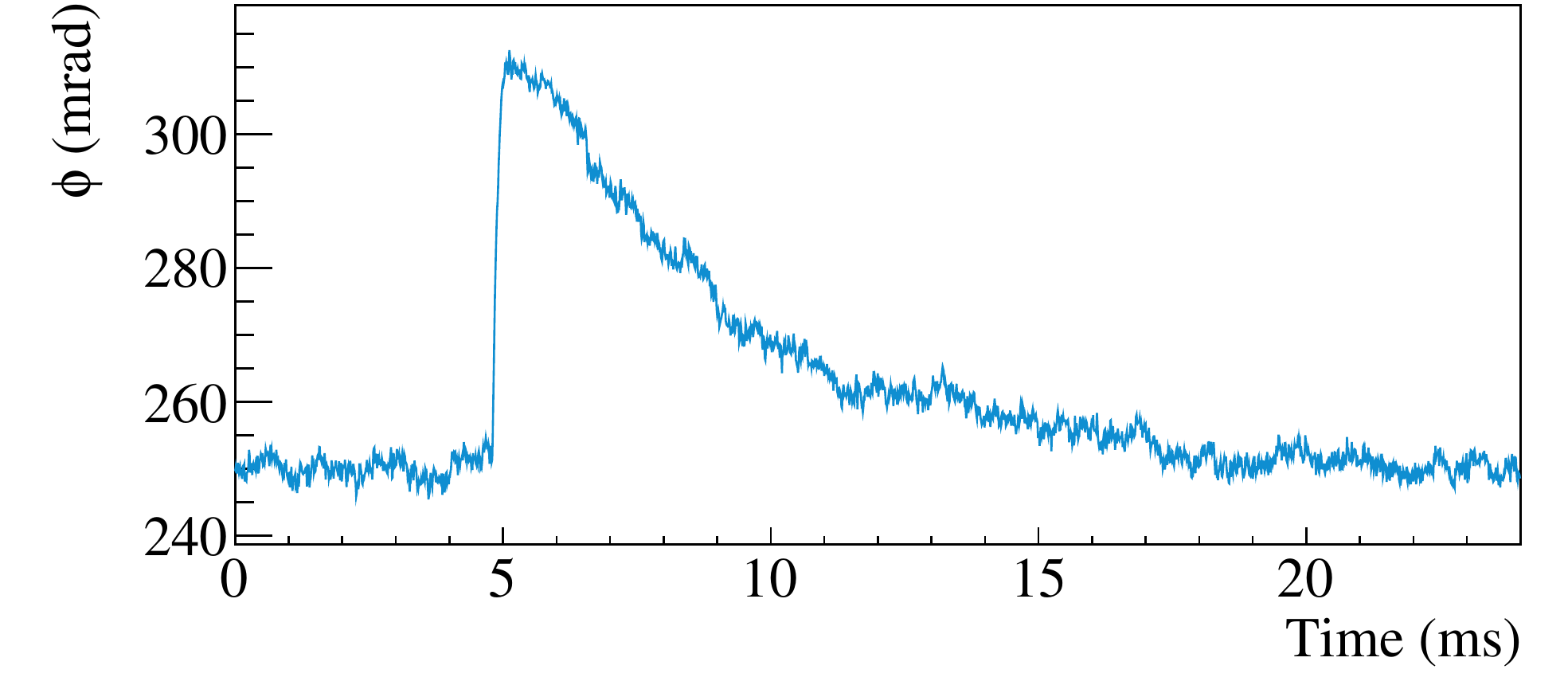}
\caption{
Raw phase pulse of KID-40 following an energy release of around 1~keV in the corresponding dice.The rise and decay times are $170~\mu$s and 3.9~ms, respectively.
}
\label{fig3}
\end{center}
\end{figure}

We choose to face the 8 optical fibers available to the KIDs marked in Fig.~\ref{fig2}c,  forming a compact cluster with $Q$ value rather homogeneous and close to the design. We restrict the data analysis only to them. Figure~\ref{fig3} shows a typical phase pulse following an energy release of $\sim1$~keV in the dice of KID-40. The rise time, computed as the time span between the 10\% and the 90\% of the rising edge, amounts to 170~$\mu s$, similar to the value expected from the resonator ring time $\log(9)\cdot Q / \pi f_0=140~\mu s$. The decay time, 
computed as the difference between the 90\% and the 30\% of the trailing edge, amounts to 3.9~ms. By monitoring the decay time at increasing cryostat temperatures we observe a significant decrease, down to 0.2~ms at 300~mK, indicating that it is generated by the recombination of quasiparticles after the pair breaking~\cite{BarendsTau}.

For the estimation of  $d\phi/dE$ we exploit the statistics of the number of photons $N$ absorbed by the dices, exploiting the technique already proved in Refs.~\cite{Cardani_2018,Cardani:2021wl}. We shine hundreds of fast ($\sim\mu s$) photon bursts at increasing LED powers and acquire the pulses triggered from the KID data stream, which are processed offline with a matched filter in order to maximize the energy resolution~\cite{Radeka:1966}. For each set of bursts the measured pulse amplitude is expected to be distributed as a Gaussian with mean $\mu= d\phi/dE\cdot\epsilon N$, where $\epsilon=3.1$~eV is the (fixed) energy of a 400 nm photon, and variance due to  Poisson fluctuations equal to $d\phi/dE\cdot\epsilon\mu$. Considering also the  resolution at zero energy $\sigma_0$ due to the KID noise, the total variance can be written as:
\begin{equation}\label{eq:poisson}
\sigma^2(\mu) =  \sigma_{0}^2 + \frac{d\phi}{dE} \epsilon \mu 
\end{equation}
The above equation is used to fit the data for  ${d\phi}/{dE}$ in order to perform the energy calibration.

We perform the photon calibration at different powers of the KID bias. Indeed, when the power is increased  the signal to noise ratio improves until the KID enter in a non-linear regime~\cite{swenson2013}. In this regime, however, the detector response is complicated by the non-linearity and $d\phi/dE$ cannot be related anymore to $\eta$ trough Eq.~\ref{eq:response}.  
Therefore  for best energy resolutions we set the power at the beginning of the non-linear regime, which corresponds to $P_{H}=-74\div-68$~dBm depending on the KID, while for the estimation of $\eta$ we lower the power to the minimum of the  electronics, $P_{L}=-86$~dBm, which is well below the non-linear region.

\begin{figure}[t]
\begin{center}
\includegraphics[width=\columnwidth]{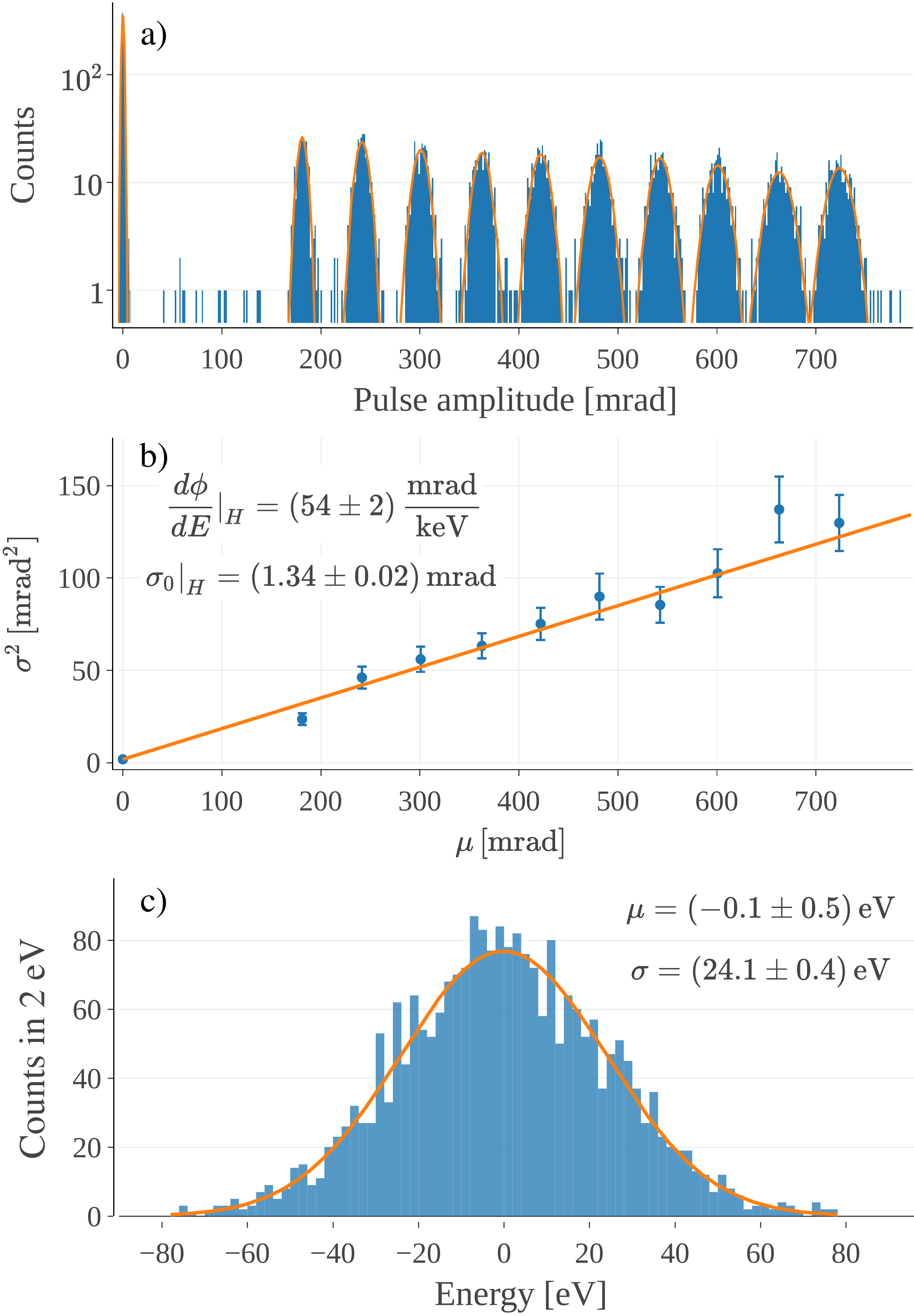}
\caption{
{\bf a)}  Measured amplitude of the phase pulses at increasing light burst intensity on the dice of KID-40. The lines show the best fits for Gaussian distributions;
{\bf b)} Variance $\sigma^2$ as a function of mean value $\mu$ of the Gaussians. {The line is the best fit for Eq.~\ref{eq:poisson}};
{\bf c)} Distribution of the calibrated noise fluctuations fitted with a Gaussian (line).
}
\label{fig4}
\end{center}
\end{figure}

Figure~\ref{fig4}a shows the amplitude  of the phase pulses of KID-40 with the bias set at $P_{H}$ for different light intensities, along with Gaussian fits to extract $\sigma$ and $\mu$ of the distributions. The point at zero amplitude is obtained from traces not containing pulses. From the fit to  Eq.~\ref{eq:poisson} of the $\sigma^2$ vs $\mu$ scattering we estimate a responsivity $d\phi/dE|_{H}=54\pm2$~mrad/keV (Fig.\ref{fig4}b). Figure~\ref{fig4}c shows the distribution of the calibrated noise, from which we can estimate a  resolution at zero energy of $\sigma_0=24.1\pm0.4$~eV. 
By performing the same operation on the other  KIDs we obtain similar results, with  average values over the 8 selected KIDs of $d\phi/dE|_{H}=43\pm9$~mrad/keV and $\sigma_0=26\pm 7$~eV.
%where the errors are the standard deviations of the distributions.
It has to be noticed that decoupling the detector from vibrations is sometimes needed in order to reach the best noise conditions of phonon sensors. 
However, our results are obtained without a decoupling system and do not change when switching the pulse tube cryocooler on and off.

When switching the power to  $P_L$, we measure on average $d\phi/dE|_{L}=57\pm 7$~mrad/keV.
For cross-check, we replaced the LED lamp with one providing photons of 255~nm instead of 400~nm and we obtained consistent results.
By inverting Eq.~\ref{eq:response} we derive $\eta=(24\pm 4)$\%, a factor 2 larger value than in the devices of CALDER. 
%{\color{red} The high value of the efficiency also indicates that the procedure we developed  for the wafer carving does not affect significantly the phonon propagation.}  
%Considering a pair-breaking efficiency of 57\%~\cite{PB}, the amount of phonons absorbed is around 40\%.
%

The high value of the energy conversion efficiency was predicted by the phonon simulations~\cite{martinez2019} we performed to design the device. Indeed, the volume of a BULLKID dice and of a CALDER substrate are similar, 146 vs 152~mm$^3$,  however in a cubic shape the average distance of a point in the volume from the sensor on a face is significantly smaller, thus enhancing  the phonon capture.
Simulations indicate also that the efficiency in a dice is limited by the thickness of the common disk which acts as a channel with adjacent dices where phonons can leak. 
We evaluate this leak by shining light on a dice and by measuring the energy detected in nearby dices. 
With respect to the energy measured in the dice of origin, we measure  on average $(14\pm 3)$\% of energy  in each of the first neighbours lying on the same line (top-bottom and left-right), and  $(5\pm1)$\% of energy in first neighbours in the diagonal direction.
These values agree with the simulations which predict also that the efficiency could be increased and the leak lowered by reducing the 0.5~mm thickness of the common disk. 

%%%%%%%%%%%%%%%%%%%%%%%%%%%%%%%%%%%%%%%%%%%%%%%
%conclusions
%%%%%%%%%%%%%%%%%%%%%%%%%%%%%%%%%%%%%%%%%%%%%%%
In conclusion our results  demonstrate the feasibility of a new concept of array of phonon detectors, consisting of a monolithic and segmented array of 60 dices of 0.34~g each sensed by multiplexed kinetic inductance detectors. In particular, the phonon energy conversion efficiency in the KID outweighs the values achieved in other detectors and pushes the energy resolution down to a few tens of eV.
%{\color{red}  Even if the analysis shown is limited to 8 units, one can assume that in terms of energy resolution the KIDs with $Q>10^5$,  corresponding to the 55\% of the array, perform similarly and this was indeed confirmed by another test in which we placed the fibers in different positions.}
The uniformity of the electrical response needs improvement, which will be achieved by improving the grounding of the feedline of the array.
Further significant improvements of the energy resolution are feasible, either by replacing the KIDs superconductor with aluminum-titanium multilayers, which already proved a factor 3 improvement with respect to aluminum, or by improving the phonon focusing in the KID with deeper carvings. Finally, once the structure of a single array is defined, tens of identical arrays could be readly produced and piled-up, giving the  opportunity to create a kg-scale, low-threshold and low-background Dark Matter or neutrino scattering experiment.

\acknowledgments
This work was partially supported by the INFN through the CSN5 grant ``BULLKID'' and ``BULLKID2''. We acknowledge the support  of the PTA and Nanofab platforms for the fabrication of the devices and of the project HAMMER~\cite{hammer} for the 3D-printed copper holder.
We thank L. Minutolo for support in the development of the electronics software,  S. Manthey Corchado for contributing in the KID design and characterization, and A. Catalano, S. Di Domizio, F. Levy-Bertrand, M. Martinez and F. Pandolfi for useful discussions.
We thank A. Girardi and M. Iannone of the INFN Sezione di Roma and V. Perino of Sapienza University for technical support.

\bibliographystyle{apsrev4-1}
\bibliography{bullkid.bbl}

\end{document}